\documentclass[	
				superscriptaddress,%
				twocolumn,%
				a4paper,
				showkeys,
				endfloats
				floatfix,%
					]{revtex4-1}
\usepackage{upgreek}
\usepackage{textgreek}
\usepackage{array}
\usepackage{enumitem}
\usepackage[dvips]{graphicx}
\graphicspath{ {./images/} }
\usepackage{color}
\usepackage{graphicx}
\usepackage{amsmath}
\usepackage{bm}
\usepackage{hyperref}
\usepackage{float}
\usepackage{makeidx}
\usepackage{hyperref}
\usepackage{float}
\usepackage{booktabs}
\usepackage{color}
\usepackage{tabularx}
\usepackage{siunitx}
\usepackage{xcolor}
\usepackage{siunitx}

\usepackage{rotating}
\usepackage{comment}
\usepackage{dcolumn}
\usepackage{makeidx}
\usepackage{multirow}
\usepackage{booktabs}
\usepackage{amsfonts}
\usepackage{braket}
\usepackage{bm}
\usepackage{float}

\usepackage{hyperref}
\hypersetup{colorlinks=true, linkcolor=blue, citecolor=blue, urlcolor=blue}

\usepackage{bm} 
\usepackage{color,soul}
\usepackage{amsfonts}
\usepackage{enumitem}

\usepackage[normalem]{ulem}

\newcommand{\mode}[2]{#1$_\mathrm{#2}$}

\newcommand{\UniLu}{University of Luxembourg, Department of physics and materials science, 41 rue du Brill, 4422 Belvaux, Luxembourg}
\newcommand{\LIST}{Materials Research and Technology Department, Luxembourg Institute of Science and Technology (LIST), 41 rue du Brill, 4422 Belvaux, Luxembourg}
\newcommand{\ULiege}{Quantum Materials Center (Q-MAT), Complex and Entangled Systems from Atoms to Materials (CESAM), Universit\'e de Li\`ege, Quartier Agora, All\'ee du Six Aout, B-4000 Li\`ege, Belgique}
\newcommand{\Neel}{Universit\'e Grenoble Alpes, Institut N\'eel CNRS, 25 Rue des Martyrs, 38042, Grenoble, France.}
\newcommand{\Soleil}{Synchrotron SOLEIL, L'Orme des merisiers, Saint-Aubin, Gif-sur-Yvette, France}

\newcommand{\ILM}{Institut Lumi\`ere Mati\`ere (ILM)-UMR 5306, Universit\'e Claude Bernard Lyon 1, Campus LyonTech-La Doua, 10 rue Ada Byron, F-69622 Villeurbanne, France}
\newcommand{\ICMCB}{Institut de Chimie de la Mati\`ere Condens\'ee de Bordeaux (ICMCB)-UMR 5026, CNRS, Universit\'e de Bordeaux, 87 Avenue du Docteur Schweitzer, F-33608 Pessac, France}

\begin{document}

\title{Stability of the tetragonal phase of BaZrO$_3$ under high pressure}

\author{Constance Toulouse*}
\affiliation{\UniLu}
\author{Danila Amoroso*}
\affiliation{\ULiege}
\author{Robert Oliva*}
\affiliation{\UniLu}
\author{Cong Xin}
\affiliation{\ICMCB}
\affiliation{\LIST}
\author{Pierre Bouvier}
\affiliation{\Neel}
\author{Pierre Fertey}
\affiliation{\Soleil}
\author{Philippe Veber}
\affiliation{\ICMCB}
\affiliation{\ILM}
\author{Mario Maglione}
\affiliation{\ICMCB}
\author{Philippe Ghosez}
\affiliation{\ULiege}
\author{Jens Kreisel}
\author{Mael Guennou}
\affiliation{\UniLu}

\begin{abstract}
In this paper, we revisit the high pressure behavior of BaZrO$_3$ by a combination of first-principle calculations, Raman spectroscopy and x-ray diffraction under high-pressure. We confirm experimentally the cubic-to-tetragonal transition at \SI{10}{\GPa} and find no evidence for any other phase transition up to \SI{45}{\GPa}, the highest pressures investigated, at variance with past reports. We re-investigate phase stability with density functional theory considering not only the known tetragonal ($I4/mcm$) phase but also other potential antiferrodistortive candidates. This shows that the tetragonal phase becomes progressively more stable upon increasing pressure as compared to phases with more complex tilt systems. The possibility for a second transition to another tilted phase at higher pressures, and in particular to the very common orthorhombic $Pnma$ structure, is therefore ruled out.
\end{abstract}

\maketitle


\section{Introduction}

The family of perovskite oxides is known for its large variety of structural distortions that is, in turn, crucial for their physical properties. The most common distortions are collective rotations (or tilts) of the corner-sharing oxygen octahedra that lower the crystal symmetry from a simple cubic structure, in the absence of tilt, to a lower symmetry determined by its tilt pattern. Such oxygen octahedra rotations are typically referred to as antiferrodistortive (AFD) distortions. There is a long history of studies of those tilts~\cite{Glazer2011}, their classification~\cite{Glazer_notation,Glazer1975,Howard2005}, and how they evolve with external parameters. 
In the most simple tilt systems, octahedra rotate only around one of the pseudo-cubic axes and the rotations in two adjacent layers perpendicular to the rotation axis can be either in-phase ($a^0a^0c^+$, in Glazer's notation~\cite{Glazer_notation}), producing a phononic instability at the $M$-point of the cubic Brillouin zone (BZ), or anti-phase ($a^0a^0c^-$), with the instability appearing at $R$-point~\cite{Safari_SZO}.
General rules for the evolution of tilts under hydrostatic pressure have been formulated~\cite{Samara1975,Zhong1995,Tohei2005,Angel2005,Xiang2017}: for most $A^{2+}B^{4+}$O$_3$ perovskites, it is established that the tilt angle should increase and, if starting from a cubic structure, cause a phase transition towards a low-symmetry tilted phase. This has been investigated very early on with studies of the cubic-to-tetragonal ($Pm\bar 3m\rightarrow I4/mcm$) phase transition in the classical perovskite SrTiO$_3$~\cite{Samara1975} and confirmed in many subsequent studies (Ref.~\cite{Grzechnik1997,Guennou2010,Yamanaka2018} and references therein). A similar scenario has been observed in other cubic perovskites including the oxide BaZrO$_3$~\cite{Chemarin2000,Yang2014} and the fluorite KMnF$_3$~\cite{Asbrink1996,Guennou2011}.

As much as this first phase transition under pressure is known, the behavior at higher pressures is still very unclear. It is natural to expect that the tetragonal phase observed under pressure in SrTiO$_3$, KMnF$_3$ or BaZrO$_3$ becomes unstable above some critical pressure, i.e. when the tilt angle reaches some critical value. This intuition partly originates from the observation of phase sequences with varying temperature that may display multiple transitions bridging different tilt systems. Particularly relevant examples are CaTiO$_3$~\cite{Redfern1996} and SrZrO$_3$~\cite{Howard2000}, which undergo phase sequences with increasing temperature $Pnma\rightarrow I4/mcm\rightarrow Pm\bar 3m$ and $Pnma\rightarrow Imma\rightarrow I4/mcm\rightarrow Pm\bar 3m$ respectively. In both cases, the cubic-to-tetragonal transition is continuous and involves a single tilt angle whereas the tetragonal-to-orthorhombic transition is discontinuous. Even richer sequences may be found in perovskites with competing polar instabilities such as NaNbO$_3$~\cite{Mishra2007, Johnson2010}. Since pressure induces much stronger reduction in bond length than temperature, we might expect \textit{cubic} $\rightarrow$ \textit{tetragonal} $\rightarrow$ \textit{orthorhombic} phase sequences for BaZrO$_3$, SrTiO$_3$ and similar perovskites under pressure. 

BaZrO$_3$ is a particularly relevant model system to study the evolution of tilts under pressure. It is experimentally cubic at ambient pressure and down to \SI{0}{\K}, but hosts a tiny instability associated to AFD distortions revealed by density functional theory (DFT) that has been attracting a great deal of attention from a theoretical point of view ~\cite{Akbarzadeh2005, Bennet2006, Bilic2009, Granhed2020, Lebedev2013}. This stirred discussions about its true ground state and the possible consequences of this instability on physical properties, such as the existence of locally distorted nanodomains, which were recently observed by electron diffraction and pair distribution function~\cite{Levin2021} or the anomaly of its dielectric constant~\cite{Akbarzadeh2005,Bennet2006}. Besides, unlike in SrTiO$_3$ and CaTiO$_3$~\cite{amoroso2018,amoroso2019} where the polar instability at the zone center plays a major role, only the zone boundary tilt mode is unstable in BaZrO$_3$ which makes it a comparatively ``pure" tilt system. Finally it has been shown that several AFD phases are nearly degenerate at ambient pressure~\cite{Toulouse_2019}, specifically the tetragonal $I4/mcm$ ($a^0a^0c^-$ in Glazer notation), the orthorhombic $Imma$ ($a^-b^0a^-$) and the rhombohedral $R\bar 3c$ ($a^-a^-a^-$). This situation leaves the energy landscape particularly open for the stabilization of multiple phases and the existence of multiple phase transitions under pressure.

Experimentally, it was indeed proposed in several instances that a second phase transition occurs both in SrTiO$_3$ and BaZrO$_3$~\cite{Yamanaka2018,Grzechnik1997,Chemarin2000,Gim2022}, but the experimental results remain controversial. In SrTiO$_3$, the existence of a second phase transition has been hypothesized based on the observation of a Raman peak splitting ~\cite{Grzechnik1997}, or changes in the Ti pre-edge features in x-ray absorption measurements~\cite{Cabaret2007,Fischer1990}. However, this proposition was discarded by Raman spectroscopy and single-crystal x-ray diffraction measurements~\cite{Guennou2010}. In BaZrO$_3$, splitting of Raman peaks was also observed and invoked to propose a transition to an orthorhombic phase, also inspired from phases found in the phase diagram of the (Ba,Ce)ZrO$_3$ system.~\cite{Chemarin2000}, and a recent Raman study concluded for a $Pm\bar 3m\rightarrow R\bar 3c\rightarrow I4/mcm$ phase sequence~\cite{Gim2022}. Nevertheless, this was not confirmed by the high-pressure x-ray study in Ref.~\cite{Yang2014}, where authors report evidence of a \textit{cubic} $\rightarrow$ \textit{tetragonal} phase transition at about $17$~GPa and persistence of the tetragonal phase up to $\sim 46$~GPa at room temperature. At the theoretical level, we are not aware of studies of the energy competition between different AFD phases in BaZrO$_3$; only predictions of pressure-induced softening of polar and antiferrodistortive modes have been proposed in Ref.~\cite{theo_zhu_2009}, as also reported for other perovskite systems~\cite{eric_BTO_2006,Kornev2007}. Also, the possibility for transitions to completely different polymorphs has been proposed~\cite{Rahmatizad2021,post_post_Perovskite}; in particular, a transition to a post-perovskite phase was predicted to occur at pressures as low as \SI{70}{\GPa} and cause a first-order transition into a semiconducting orthorhombic phase with a 7.9$\%$ volume collapse. These predictions have not yet found experimental confirmation.

In this paper we revisit the behavior of BaZrO$_3$ under hydrostatic pressure with the aim to clarify its pressure-induced phase sequence. We show experimental data by Raman spectroscopy and x-ray diffraction (XRD), and discuss their interpretations also relying on supporting first-principle calculations of the dynamical properties and energetics as a function of pressure. 
The similar case of SrTiO$_3$ is also discussed for comparison. 
Our study provides strong arguments for excluding other structural transitions than the cubic-to-tetragonal up to $45$~GPa, and provides explanations for the divergent claims found in the literature.

\section{Methods and experimental details}

Single crystal samples were prepared from a crystal purchased from Crystal~Base~Co.~Ltd. and grown by the tri-arc Czochralski method, as described and characterized in Ref.~\cite{Xin2019}.

We performed single crystal Raman measurements under pressure up to \SI{20}{\GPa} in a membrane-type diamond anvil-cell (DAC) with a \SI{600}{\um} culet size using a 4:1 methanol-ethanol solution as pressure-transmitting medium. Pressure was measured in situ with two ruby balls inserted in the chamber together with the BaZrO$_3$ single crystal sample. The two ruby were giving the same pressure up to the highest pressures showing good hydrostaticity. The spectra were recorded using a \SI{532}{\nm} laser line and a Renishaw inVia confocal Raman spectrometer calibrated using the \SI{521}{\per\cm} phonon mode of Silicon.

Two sets of powder x-ray measurements were performed at the ID27 beamline of the ESRF synchrotron facility. Both measurements were performed using Neon as a pressure transmitting medium. A monochromatized x-ray radiation source of wavelength \SI{0.3738}{\angstrom} and a two dimensional charge-coupled device detector (MAR-CCD) with a pixel size of \SI{79}{\um} were used. A Silicon powder standard was used to calibrate the distance of the detector and other geometry parameters. For the first set, powdered BaZrO$_3$, refered to here as sample S1, was loaded on a DAC with a culet size of \SI{300}{\um} and diffraction patterns up to \SI{16.5}{\GPa} with pressure steps of \SI{0.7}{\GPa} were obtained. The second set of measurements were performed up to \SI{45}{\GPa} using a DAC with a culet size of \SI{250}{\um} and a second powder sample of BaZrO$_3$, called S2, prepared as before. The pressure was determined from the ruby calibration method~\cite{ruby_method} and the standard deviation in pressure measurement was of less than 1\%, ensuring a good hydrostaticity~\cite{Klotz_2009}. The data was analyzed by full Rietveld refinements using FULLPROF software~\cite{Rodriguez1993}. The refined parameters are; $i)$ the scale factor, $ii)$ manually chosen linearly interpolated background at 15 points, $iii)$ lattice parameters $a$ and $c$, $iv)$ fractional coordinates, $v)$ isotropic thermal parameters and $vi)$ shape and broadening parameters. The peak shapes were described with a pseudo-Voigt function. The profile parameters $u$, $v$, $w$, $lx$ and $ly$, which determine the resolution function (see FULLPROF manual), were obtained from the refinement of a high-purity silicon standard.

For single crystal XRD measurements, two small crystals with different orientations were loaded into a DAC and gave essentially identical results; the corresponding data will labelled as sample S3 in the following. Measurements were performed at the CRISTAL beamline of the SOLEIL synchrotron facility. A monochromatized x-ray radiation source of wavelength of \SI{0.41579}{\angstrom} and a two dimensional charge-coupled device (2D CCD, Rayonix SX165) with a pixel size of \SI{79}{\um} and a diameter of \SI{165}{\mm} were used. A ruby single crystal and Lanthanum hexaboride (LaB$_{6}$) powder were used to calibrate the geometrical parameters of the diffractometer and the wavelength respectively. The pressure inside the \SI{600}{\um} culet-sized DAC was determined from the diffraction of powdered gold included near the samples and its reported equation of state (EoS)~\cite{gold_EoS,Dewaele2004}. Helium was used a pressure-transmitting medium. The experiment was conducted up to \SI{12.5}{\GPa} at which point the diamonds got damaged, impeding to carry out a full structural refinement at higher pressures.

First-principle calculations of structural and dynamical properties rely on DFT and density functional perturbation theory (DFPT)\cite{dfpt_94,dfpt_97}, as implemented in the ABINIT package~\cite{abinit1,abinit2,gonze2020}. The exchange-correlation potential was evaluated within the generalized gradient approximation (GGA) using the Wu-Cohen (WC) functional~\cite{wc-gga}, which provides us with good description of the structural properties of the cubic phase and a lattice parameter $a_0$ of about \SI{3.184}{\angstrom} -- cf. experimental $a_0\simeq3.195$~\AA~(Fig.~\ref{fig:xrd_all}.c) and lattice dynamics reported in our previous work~\cite{Toulouse_2019} -- correcting the reported overestimation within the more conventional Perdew-Burke-Ernzerhof (PBE) functional~\cite{Granhed2020} or underestimation within the Local Density Approximation (LDA) functional~\cite{theo_zhu_2009,eric_BTO_2006}. Norm-conserving pseudopotentials~\cite{pseudo} have been employed with the following orbitals considered as the valence states: $5s$, $5p$, and $6s$ for Ba $4s$, $4p$, $4d$, and $5s$ for Zr, and $2s$ and $2p$ for O. The energy cutoff for the expansion of the electronic wave functions has been fixed to 45 Ha and we used a 6$\times$6$\times$6 k-point mesh for the Brillouin zone sampling. Phonon calculations have been performed on the fully relaxed cubic and AFD structures.
In particular, structural optimization to find the equilibrium configuration of the ions and lattice parameters were performed using the Broyden-Fletcher-Goldfarb-Shanno minimization (BFGS); the maximal absolute force tolerance was fixed to $10^{-5}$ Ha/Bohr. A negative and isotropic target stress tensor was imposed during the geometry optimization to simulate the hydrostatic pressure effect.
The phonon dispersion curves for the 5-atom cubic cell of BaZrO$_3$ have been obtained through Fourier-based interpolation of the dynamical matrices, as implemented in the post-processing tool related to the ABINIT package, ANADDB~\cite{gonze2020}. Direct computation of the full dynamical matrices was performed via DFPT at the $\Gamma$, $X$, $M$, $R$ and the $\Lambda$ (halfway from $\Gamma$ to $R$) points of the simple cubic Brillouin zone.

For the comparison with the Raman data and analysis, phonon frequencies of the AFD structures have been calculated at the $\Gamma$ point. Symmetry analysis of the phonon modes was done with the help of programs from the Bilbao crystallographic server~\cite{kroumova-bilbao-2003,elcoro-bilbao-2017} and the ISOTROPY Software Suite~\cite{isotropy}.
 
\section{Results and discussion}

\subsection{Raman spectroscopy}
\label{sec:raman}
\begin{figure*}[t!]
\resizebox{\textwidth}{!}{\includegraphics{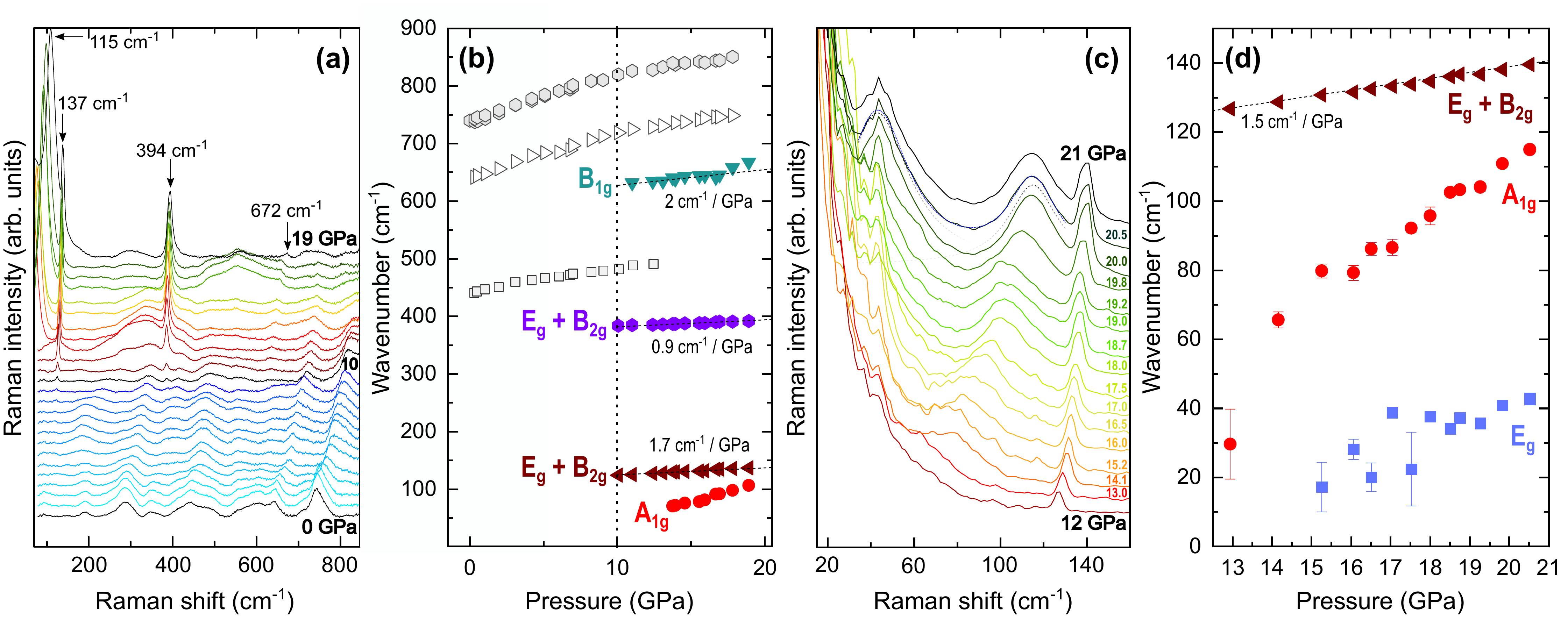}}
\caption{Raman spectra of BaZrO$_3$ single crystals under pressure (a) from 0 to \SI{19}{\GPa}, when increasing the pressure, and (c) between 21 and \SI{12}{\GPa}, when releasing the pressure. The spectra in (a) are classical Stokes spectra, while the spectra in (c) are collected without cutting off the Rayleigh scattering of the laser line, allowing for observation of lower energies. An example of the Lorentzian fits performed on the Raman modes is shown in blue. (b) Pressure dependence of the wavenumbers of the phonon modes derived from the spectra obtained from 0 to \SI{19}{\GPa}, the solid symbols show the modes appearing above the transition at \SI{10}{\GPa}. (d) Behaviour of the two soft modes (and first hard mode) appearing above \SI{10}{\GPa}, these frequencies are derived from the spectra shown in (c), fitted with Lorentzian functions.}
\label{fig:Raman}
\end{figure*}

\begin{table*}[th!]
	\centering
	\caption{Wavenumbers (in cm$^{-1}$ units) at \SI{20}{\GPa} of the Raman-active phonon modes calculated through DFPT for the three tilted phases considered here as well as values measured experimentally. The full table including Raman inactive modes is included in the supplementary material. The modes are grouped together based on the mode they originate from in the cubic phase. 
	}
	\label{tab:DFT_modes_freq}
	\setlength{\extrarowheight}{2pt} 
	\begin{tabularx}{0.95\textwidth}{c|l|>{\centering}m{0.06\textwidth}>{\centering}m{0.06\textwidth}|>{\centering}m{0.07\textwidth}>{\centering}m{0.06\textwidth}|>{\centering}m{0.06\textwidth}>{\centering}m{0.07\textwidth}|c}
		\hline\hline
		Cubic & Vibrational pattern &
		\multicolumn{2}{c|}{Tetragonal} & \multicolumn{2}{c|}{Orthorhombic} & \multicolumn{2}{c|}{Rhombohedral} & Experimental \\
		($Pm\bar{3}m$ - 221) & &
		\multicolumn{2}{c|}{($I4/mcm$ - 140)} & \multicolumn{2}{c|}{($Imma$ - 74)} & \multicolumn{2}{c|}{($R\bar{3}c$ - 167)} & Frequencies\\
        \hline
		\multirow{3}{*}{R$_4^+$} & & 76 & \mode{E}{g} & 39 & \mode{B}{2g} & 23 & \mode{E}{g} & 48 \\
		 & oxygen octahedra rotations & & & 69 & \mode{B}{1g} & & & \\
		 & & 189 & \mode{A}{1g} & 182 & \mode{A}{g} & 177 & \mode{A}{1g} & 115 \\
		\hline
		\multirow{3}{*}{R$_5^+$} & & 138 & \mode{E}{g} & 135 & \mode{B}{2g} & 142 & \mode{E}{g} &\\
		 & antiparallel Ba motion & & & 134 & \mode{B}{3g} & & & 137\\
		 & & 145 & \mode{B}{2g} & 139 & \mode{A}{g} & & & \\
		\hline
        \multirow{3}{*}{R$_5^+$} & oxygen octahedra & 374 & \mode{E}{g} & 363 & \mode{B}{2g} & 383 & \mode{E}{g} & \\
		 & shearing modes & & & 373 & \mode{B}{3g} & & & 394\\
		 & & 385 & \mode{B}{2g} & 390 & \mode{A}{g} & & & \\
		\hline
        R$_3^+$  & Jahn-Teller-like distortions & & & 662 & \mode{B}{1g} & 665 & \mode{E}{g} & \multirow{2}{*}{672}\\
		 & of oxygen octahedra & 665 & \mode{B}{1g} & 665 & \mode{B}{3g} & & & \\
		\hline
		R$_1^+$ & oxygen octahedra breathing mode & & & 896 & \mode{B}{3g} & & & \\
		\hline
		\hline
	\end{tabularx}
\end{table*}

The Raman spectra collected under pressure up to \SI{21}{\GPa} are shown in Fig.~\ref{fig:Raman}.a and \ref{fig:Raman}.c. At ambient pressure, in the cubic phase, BaZrO$_3$ has no Raman active phonon modes but nonetheless exhibits an intense Raman spectrum consisting in broad second-order bands as described in Ref.~\cite{Toulouse_2019} and references therein. Upon increasing pressure, this second-order spectrum weakens. At \SI{10}{\GPa}, some additional sharp peaks emerge, revealing the activation of Raman modes due to the structural phase transition. In addition, a soft mode comes into view from lower frequencies with high intensity (Fig.\ref{fig:Raman}.a), followed by a second soft mode at lower wavenumbers only visible when measuring very close to the quasi-elastic line (Fig.\ref{fig:Raman}.c). The frequency evolution under pressure of the Raman peaks, fitted with Lorentzian functions, is shown in Fig.~\ref{fig:Raman}.b and \ref{fig:Raman}.d. The hard modes exhibit slopes ranging from \SI{0.9}{\per\cm\per\GPa} to \SI{2}{\per\cm\per\GPa} (see Fig.\ref{fig:Raman}.b and \ref{fig:Raman}.d). 

An assignment of the modes appearing at the transition can be done with the help of the analysis detailed in our previous paper \cite{Toulouse_2019} and partially reproduced in Table~\ref{tab:DFT_modes_freq}. The mode that appears at \SI{124}{\per\cm} is assigned to Ba motion and the mode appearing at \SI{384}{\per\cm} to oxygen octahedra shearing modes, while the soft modes are known to correspond to tilts of the octahedra. Only the totally symmetric soft mode can be followed with accuracy, the lower energy soft mode (visible in Fig.\ref{fig:Raman}.c and fitted in Fig.\ref{fig:Raman}.d) gets lost in the  quasi-elastic line at low pressures. Note that the hard mode appearing at \SI{124}{\per\cm} (at \SI{140}{\per\cm} at \SI{20}{\GPa}) has an asymmetric profile at higher pressures, which may indicate the presence of two close overlapping bands.

Our single crystal spectra are overall in agreement with the single crystal spectra from Ref.~\cite{Gim2022} and the powder spectra from Ref.~\cite{Chemarin2000}, but also differ in some aspects. The major difference is that an intense peak appears in the powder spectrum at the phase transition and reaches ~\SI{680}{\per\cm} at \SI{20}{\GPa}. On our single crystal spectra, we do observe a feature that becomes visible at \SI{634}{\per\cm} around 11-12~GPa and hardens up to \SI{670}{\per\cm} at \SI{19}{\GPa} (Fig.~\ref{fig:Raman}.a) which can correspond to the third hard mode observed in Ref.~\cite{Chemarin2000}, but with significantly less intensity. The very same mode does appear very clearly in the single crystal study by Gim et al.~\cite{Gim2022} This difference in intensity between different experiments presumably arises from orientation effects due to selection rules, as will be discussed in the following.

Next, we focus on the identification of the symmetry of the high pressure phase. In Ref.~\cite{Chemarin2000}, the first high pressure phase was identified as $R\bar 3c$ by analogy with the Raman spectrum of Ba$_{0.9}$Ce$_{0.1}$ZrO$_3$, and because the Raman modes appeared first as single peaks. It was then proposed that a second phase transition occurs at higher pressures based on the splitting of the hard Raman modes located around 150 and \SI{380}{\per\cm}. The same reasoning was made in Ref.~\cite{Gim2022}, but the second high-pressure phase was identified as tetragonal instead. As stated in our previous study, identifying the high pressure structure from Raman spectra may not be as easy an anticipated. This is because, even though the number of Raman active modes is theoretically different, they all originate from the same degenerate modes at the $R$ point of the cubic phase, so that differences between the tetragonal, orthorhombic and rhombohedral structures are only revealed by peak splittings that scale with the very small distortion of the unit cell, and mode polarizations that are difficult to appreciate without performing polarized experiments. In such a context, identifying the correct number of Raman active modes can be troublesome.

In order to properly identify the high-pressure phase, we provide in Table~\ref{tab:DFT_modes_freq} the wavenumbers of the Raman active modes computed by DFPT at \SI{20}{\GPa} for the three considered phases with antiphase tilts (tetragonal $I4/mcm$, orthorhombic $Imma$ and rhombohedral $R\bar 3c$) together with our experimental Raman values at \SI{20}{\GPa}. From this table, we make the following observations. First, the orthorhombic phase is markedly different from the two others mainly by three criteria; \emph{i)} splitting of the soft mode into three modes instead of two, \emph{ii)} large splitting of the second hard mode with a totally symmetric mode given here at \SI{390}{\per\cm}, and \emph{iii)} Raman activation of the octahedron breathing mode at \SI{896}{\per\cm}. Octahedra breathing modes in perovskites usually have strong intensities when symmetry allowed, but here they unfortunately overlap with very strong bands from the pressure transmitting medium and nothing can be said about this mode neither in our work nor in Refs.~\cite{Chemarin2000,Gim2022}. On the other hand, none of these three Raman studies shows a peak splitting of the second hard mode. It appears very unlikely that such a splitting could occur without being detected, since the modes all originate from the same atomic displacement patterns and would have {\it a priori} comparable Raman intensities, especially in a powder average. We therefore exclude here the possibility of an orthorhombic phase. Distinguishing between the rhombohedral and the tetragonal phase is more delicate. They differ only in mode symmetries and with very moderate peak splitting (smaller than \SI{10}{\per\cm}), which can be easily overlooked given the observed peak widths. A detailed examination of mode symmetries strongly supports polarization effects as the origin for the differences observed between the powder spectra from Ref.~\cite{Chemarin2000} and our single crystal spectra. Hard modes split into modes of different symmetries, which is more easily observed with the orientational average of a powder than on single crystals. Interestingly, the hard mode appearing at \SI{672}{\per\cm} is predicted to have a \mode{B}{1g} symmetry in the tetragonal phase, which makes it particularly difficult to observe on single crystal measurements if the tetragonal $c$ axis is in the plane of the DAC. In contrast, it has the same symmetry (\mode{E}{g}) as the other hard modes in the rhombohedral phase, and no particular reason to be less visible. Therefore, comparison between our single crystal and the powder spectra from Ref.~\cite{Chemarin2000} points towards the tetragonal phase as the most probable. This assignation becomes in turn perfectly compatible with the peak splitting reported at higher pressures: at low pressures close to the transitions, peak splitting is very weak and these doublets \mode{E}{g}$\oplus$\mode{B}{2g} appear as a single peak; the splitting is resolved only at higher pressures when the distortion become sizeable (at \SI{30}{\GPa} in Chemarin et al.~\cite{Chemarin2000} and \SI{19.2}{\GPa} in Gim et al.). With this reasoning, we do agree with Gim et al.~\cite{Gim2022} on their assignement of the high pressure phase to the tetragonal $I4/mcm$ variant, but disagree on the presence of an intermediate rhombohedral phase which is not necessary to explain the observations, and also would lead to first-order transition that is not compatible with the smooth evolutions observed in all studies. In summary, Raman spectroscopy supports the hypothesis of a single phase transition towards a tetragonal $I4/mcm$ phase and no further transition up to the maximum pressure of \SI{42}{\GPa} reached by Gim et al. 

The exact same discussion can be made about SrTiO$_3$ and the conflicts on its Raman spectrum, with the additional simplicity that literature agrees on a single tetragonal phase. The splittings of the hard modes at 185 and \SI{470}{\per\cm} reported in Ref.~\cite{Grzechnik1997} and interpreted as a possible phase transition towards an orthorhombic phase remain in fact compatible with a tetragonal symmetry. Also, the mode around \SI{600}{\per\cm} is seen less clearly on a single crystal~\cite{Guennou2010} than on powder~\cite{Yamanaka2018}. Even though this mode is not clearly assigned in those past studies, it is now clear that it follows the behavior and polarization effects described above and in Table~\ref{tab:DFT_modes_freq}. We also note that the Raman signatures (gradual vanishing of the second-order Raman bands, emergence of the \mode{E}{g} soft mode) are essentially identical to the signatures reported for SrTiO$_3$ in Ref.~\cite{Guennou2010} and are all compatible with the tetragonal structure. Altogether, SrTiO$_3$ and BaZrO$_3$ appear as perfectly isostructural in their high-pressure behavior from the Raman point of view.

\subsection{X-ray diffraction}

\begin{figure*}[]
\begin{center}
\includegraphics[width=0.9\textwidth]{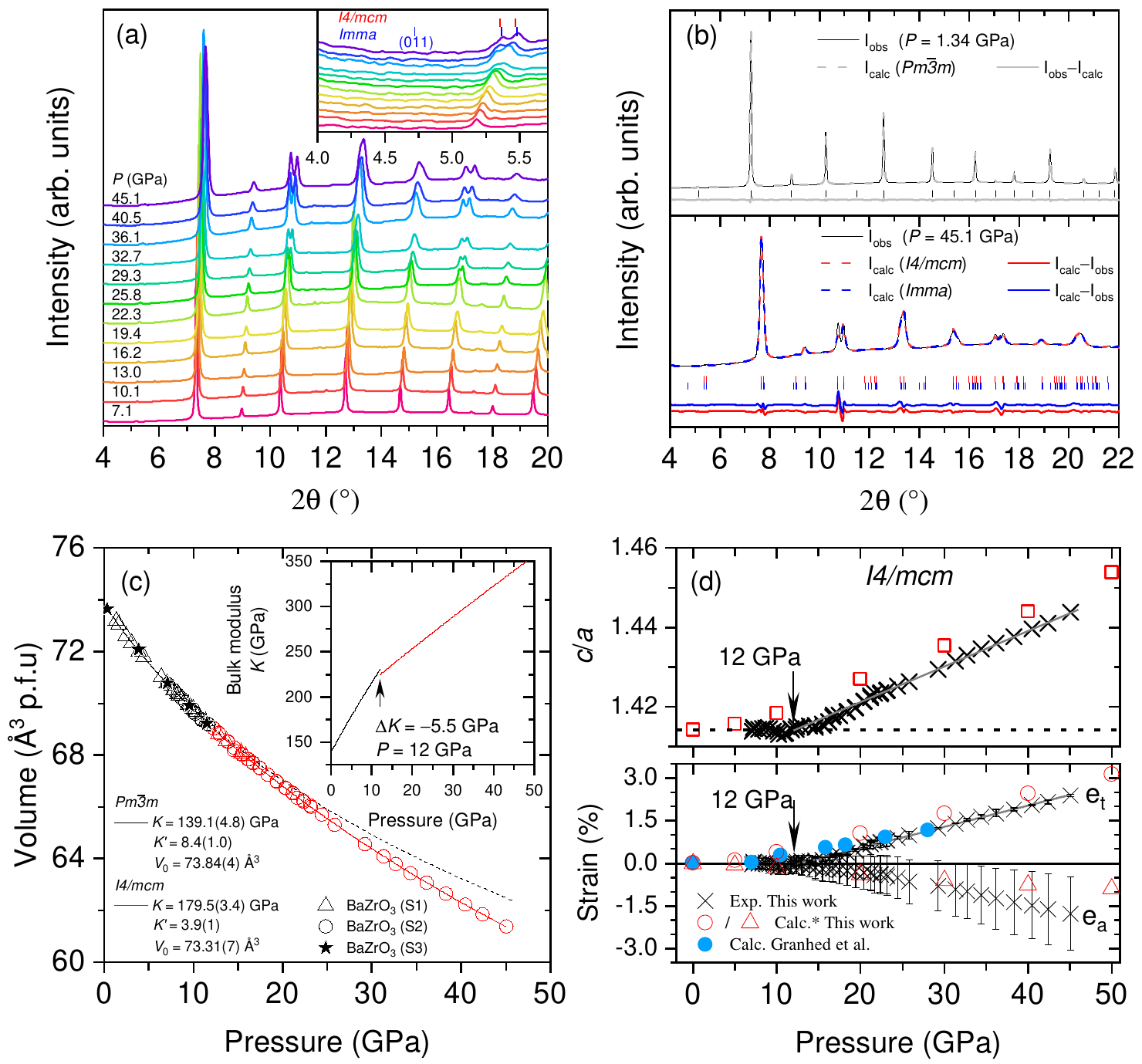}
\caption{{\bf (a)} Diffraction patterns of sample S2. Inset: Zoom of the low-angle region. Low-angle fitted peak positions at \SI{45.1}{\GPa} are included as red and blue ticks corresponding to $I4/mcm$ and $Imma$ structures, respectively. Note that no diffraction peak is visible where the reflection (011) of an $Imma$ phase would be expected. {\bf (b)} Top: Rietveld fit (grey line) to a diffraction pattern (black line) acquired at \SI{1.34}{\GPa}. Bottom: Rietveld fit assuming tetragonal and orthorhombic phases (red and blue dashed lines, respectively) to a pattern acquired at a pressure of \SI{45.1}{\GPa}. Fitted positions for all reflections are included at the bottom as colored ticks. Both panels use the same intensity scale and residuals of the $Imma$ fit are vertically shifted for a better comparison. {\bf (c)} Pressure dependence of the volume per formula unit ABO$_3$ for the cubic and tetragonal phases (black and red, respectively).  Lines are fits to BM EoS (for the cubic phase a high pressure extrapolation is included as a black dashed line). Inset: Bulk modulus as a function of pressure. {\bf (d)} Top: Tetragonality ($c/a$ ratio) as a function of pressure as obtained from the fitted x-ray diffraction (XRD) patterns of sample S2 (crosses) and adapted DFT calculations (squares). Bottom: Volume ($e_a$) and tetragonal ($e_t)$ spontaneous strains as obtained from the experiments (crosses), our calculations (open symbols) and calculations published by Granhed et al.~\cite{Granhed2020}. Our reported DFT-values include a pressure shift (+10 GPa) as discussed in the text; this is not the case for values taken from Ref.~\cite{Granhed2020} calculated through the hybrid Heyd-Scuseria-Ernzerhof (HSE) functional, which provides a slight larger volume than the one obtained from the WC-functional employed here (cf. Table~I in Ref.~\cite{Granhed2020}).} 
\label{fig:xrd_all}
\end{center}
\end{figure*}

Diffraction patterns measured for sample S2 are shown in Fig.~\ref{fig:xrd_all}.a up to a pressure of \SI{45.1}{\GPa}. A splitting of the peaks corresponding to the pseudo-cubic (200)$_\mathrm{pc}$ and (310)$_\mathrm{pc}$ reflections around 10.6$^{\circ}$ and 17$^{\circ}$ respectively is particularly visible from a pressure of \SI{25.8}{\GPa}. Other pseudo-cubic reflections show broadening with potentially unresolved peak splitting. The rhombohedral structure is not compatible with the splitting of the (200)$_\mathrm{pc}$ reflection and can therefore be safely ruled out.  (See also Fig.~S5 in the SM for more details). On the other hand, distinguishing between the tetragonal and orthorhombic structures may not be trivial if the orthorhombic distortion is small and the metric quasi-tetragonal. One possible criterion is to check the low-angle region where a (011)$_\mathrm o$ reflection is expected for the orthorhombic phase only as a result of the off-centering of the Ba cation. The inset in Fig.~\ref{fig:xrd_all}.a shows an enlarged view of this region where no reflection can be seen even at the highest pressures. From Rietveld simulations for the $Imma$ structure and with our level of noise, this indicates that Ba displacement cannot be larger than $\approx 0.007$, a value that is small but not unrealistically; small when comparing to values found in our DFT results. This criterion alone is therefore not sufficient to conclusively rule out the orthorhombic phase. 

Another criterion is to examine the values of the (pseudo)tetragonality, based on the following observation. Both the $I4/mcm$ and the $Imma$ unit cells are similar, with one lattice vector that is doubled with respect to a primitive cubic lattice vector. In the $I4/mcm$ phase, the octahedra rotation axis is along this doubled axis, which is then elongated while the perpendicular axes shrink, resulting in a tetragonality $c/a$ larger than $\sqrt{2}$. In contrast, in the $Imma$ structure, the rotation axis is perpendicular to the doubled axis ($b$ in the standard setting) so that this axis shrinks and the resulting pseudotetragonality is lower than $\sqrt{2}$. This is verified experimentally notably in the $Imma\rightarrow I4/mcm$ transition in SrZrO$_3$~\cite{Howard2005} as well as in other $Imma$ perovskites~\cite{Kennedy2004,Kukuse2016}. Here, the $c/a$ ratio is most definitely larger than $\sqrt{2}$. We can therefore discard the orthorhombic phase and retain the tetragonal $I4/mcm$ as the only high-pressure phase.

Rietveld refinement of the patterns were performed in the cubic and tetragonal phases. As illustrated in the top panel of Fig.~\ref{fig:xrd_all}.b, the diffraction pattern of S1 can be perfectly fitted at \SI{1.34}{\GPa} in the cubic phase. Refinements are less satisfactory at higher pressures, as illustrated with the pattern at \SI{45.1}{\GPa} shown in the bottom panel of Fig.~\ref{fig:xrd_all}.b. This is expected due to a general degradation of the pattern quality at high pressure following inter-grain stress, loss of hydrostaticity, etc. Unfortunately, the quality of the pattern was not sufficient to determine reliably the oxygen positions, it is therefore not possible to give a value for the tilt angle itself. For completeness, we also performed a Rietveld refinement of the pattern at \SI{45.1}{\GPa} in the orthorhombic phase (also displayed in Fig.~\ref{fig:xrd_all}.b), only to note that it does not improve the quality of the refinement as compared to the tetragonal case (reliability factors are, for $Imma$, $R_{p}$ = 16.3, $R_{wp}$ = 13.0 and ${\chi}^2$ = 0.58 while slightly better values are found for $I4/mcm$ $R_{p}$ = 12.8 , $R_{wp}$ = 12.3 and ${\chi}^2$ = 0.59 due to small differences in the fitting procedure) which further validates the tetragonal assignation. 

The pressure-volume relation of S1, S2 and S3 are shown in Fig.~\ref{fig:xrd_all}.c. All data were adequately fitted using the 3$^\mathrm{rd}$-order Birch-Murnaghan (BM) EoS (a discussion on the choice of EoS is provided in Fig. S6 of the SM based on finite-strain analysis~\cite{Jeanloz1991}), fitted volumes at zero pressure as well as bulk moduli and its pressure derivative are provided in the figure. For the tetragonal phase, very small deviations (i.e. volume deviations smaller than 0.1\%, as shown in the supplementary material, SM, Fig.~S7) are found from the fitted EoS up to the highest pressure, which is another indication that no other phase transition takes place and that only the $I4/mcm$ structure remains stable at least up to \SI{45.1}{\GPa}. There is no measurable volume jump at the transition. The pressure dependence of the bulk modulus in both phases has been included as an inset to Fig.~\ref{fig:xrd_all}.d in order to better illustrate the changes of second derivative of the Gibbs energy at the transition pressure. As it can be seen in the inset, at the transition pressure the bulk modulus exhibits a decrease of slope (due to the decreased $K'$ for the high-pressure phase) together with a small variation ($\Delta K =$ \SI{-5.5}{\GPa}), which is consistent with a second-order transition.

From the Rietveld refinements we extracted the lattice parameters for both phases for samples S1 and S2. The tetragonality $c/a$ is plotted versus pressure in Fig.~\ref{fig:xrd_all}.d. 
We adopt here the symmetry-adapted spontaneous strains defined as $e_{a}=(2e_{1}+e_{3})$ (volume strain) and $e_{t}=2(e_{3}-e_{2})/\sqrt[]{3}$ (tetragonal strain), where $e_{1}$ and $e_{3}$ are the spontaneous strain components of the tetragonal phase with respect the cubic phase, $e_{1}=e_{2}=(a_{pc}-a_{0})/a_{0}$ and $e_{3}=(c_{pc}-a_{0})/a_{0}$~\cite{Carpenter2007}. We consider $a_{0}=V^{1/3}$, where $V$ is a cubic volume extrapolated from the EoS of the cubic phase whereas the pseudocubic lattice constants were calculated from the tetragonal lattice constants, $a_{pc}=a_{t}/\sqrt{2}$ and $c_{pc}=c_{t}/2$. As it can be seen in the lower panel of Fig.~\ref{fig:xrd_all}.d the tetragonal strain $e_t$ is positive and linear with pressure with a slope of 7.5$\cdot$10$^{-4}$ GPa$^{-1}$. The volume strain $e_a$ is negative and exhibits comparatively large errors due to the increasing uncertainty of the extrapolated EoS at higher pressure. 
Good agreement is found with respect to the DFT-prediction, as shown in Fig.~\ref{fig:xrd_all}.d.
Noteworthy, both $e_a$ and $e_t$ values are very similar to those reported for SrTiO$_3$~\cite{Guennou2010}.  

Overall, the combination of XRD and Raman spectroscopy data confirms that BaZrO$_3$ undergoes a phase transition from cubic to tetragonal but also show that no other transition happens up to the maximum experimental pressure investigated of \SI{45.1}{\GPa}.

The value of the transition pressure varies slightly depending on the method used. From the powder diffraction data, it can be determined by an extrapolation of the tetragonality to its ideal cubic value of $\sqrt{2}$ (considering $c_t/a_t$), or of the tetragonal strain to zero, which gives a value of \SI{12}{\GPa}. This is in line with the emergence of superstructure reflections observed in the single crystal diffraction patterns only slightly above the noise level at \SI{11.2}{\GPa} and much more conclusively at \SI{12.5}{\GPa} (Fig.~S4 in the SM). In the Raman data, the first-order peaks appear at \SI{10}{\GPa}, which is consistent with the $\approx$11 GPa measured from Raman spectroscopy in Ref.~\cite{Chemarin2000}. The slightly lower value found by Raman spectroscopy as compared to XRD is consistent with the smaller coherence length of this technique and its sensitivity to small structural changes. We note that our values are significantly lower than the \SI{17.2}{\GPa} given in Ref~\cite{Yang2014} where it was determined only from the observation of the peak splitting in a powder XRD pattern. In the following, we chose to retain the value of \SI{10}{\GPa} measured by Raman spectroscopy.

\subsection{First-principle calculations}

\begin{figure*}[t]
\begin{center}
\includegraphics[width=1\textwidth]{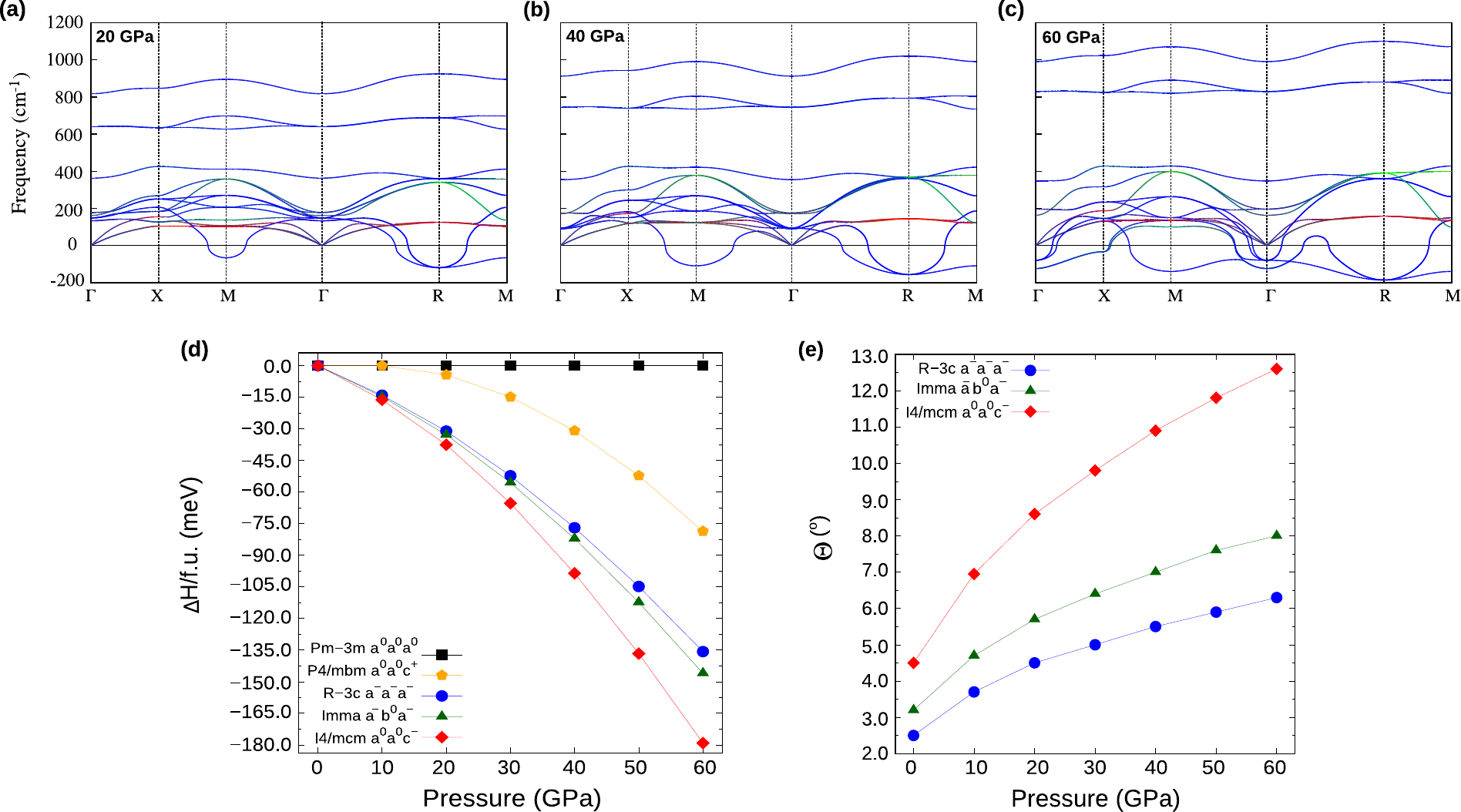}
\caption{{\bf (a,b,c)} DFPT phonon dispersion curves in the cubic phase for 20, 40 and \SI{60}{\GPa} applied hydrostatic pressure (corresponding lattice parameters $a$ are $4.049$ \AA, $3.955$~\AA~ and $3.883$~\AA, respectively). Red, green and blue phononic branches involve mostly Ba, Zr and O vibrations, respectively. {\bf (d)} Computed enthalpies (H), with respect to the cubic phase taken as reference at each pressure. {\bf (e)} Evolution of the oxygen octahedra rotation angle $\Theta$ (in degree, $^\circ$) with isotropic external pressure for different rotation patterns. }
\label{fig:dft_calc}
\end{center}
\end{figure*}

The propensity of BaZrO$_3$ to undergo a pressure-induced structural phase transition can be inferred from the inspection of phonon frequencies calculated in the compressed cubic reference structure, as well as from the DFPT-calculated phonon dispersion curves shown in Fig.~\ref{fig:dft_calc}.a-c (and Fig.~S1; cf. also Ref.~\cite{theo_zhu_2009}). As it can be seen in the figure, an increase of external hydrostatic pressure ($P$) 
results in enhanced instabilities of the high-symmetry cubic phase arising from oxygen motion. In fact, increasing $P$ enhances the AFD instabilities associated with the cooperative rotations of the oxygen octahedral cage (Fig.~\ref{fig:dft_calc}.d). For a wide range of pressure, up to \SI{40}{\GPa}, the main driving instability is the one at $R$, accompanied by that at $M$, which is its continuation in the phonon spectrum. Higher pressure destabilize the system further, making polar and antipolar instabilities appearing at the $\Gamma$ and $X$ points. 

It should be noted that, from our DFT calculations, the cubic phase is already unstable at \SI{0}{\GPa}~\cite{Toulouse_2019,Akbarzadeh2005,Bennet2006,Granhed2020,Bilic2009}, but can be made stable with a negative pressure of about $-10$~GPa (Fig.~S1). Such behaviour can be ascribed to known effects related to the used exchange-correlation functional in the calculations~\cite{perrichon_2020}. Besides, first-principles simulations do not account for temperature effects, which can strongly affect the critical pressure ($P_c$) at which the phase transition occurs, shifting it towards a higher value. 
For example, the phase diagram of SrTiO$_3$ exhibits a P-T slope of \SI{0.054}{\GPa\per\K}, which causes a shift in transition pressure of $\Updelta P_c$ = \SI{16.2}{\GPa} between 0 and \SI{300}{\K}~\cite{Guennou2010}. Here, a shift of $+10$~GPa has been added to the DFT data to match the transition point, when comparing the calculated $c/a$ ratio and spontaneous strains with the experimental ones. This is here appropriate since we are interested in trends and the energetics of the different phases rather than in the accurate and absolute values of transition pressures. 
On a related note, the thermal expansion is not strong here. In fact, the relaxed lattice parameter (by DFT) of the cubic structure at \SI{0}{\K}--\SI{0}{\GPa} ($\simeq 4.184$~\AA~) perfectly reproduces the low-temperature (\textit{T} = 4.2 K) experimental value ($\simeq 4.188$~\AA), but is also very similar to that measured at room temperature ($\simeq 4.193$~\AA)~\cite{knight2020}. A good agreement is also found for the volume and lattice parameters of the AFD structures upon external pressure, with respect to those experimentally estimated via the Rietveld refinements of the XRD data (Fig.~S3.a-b). 


Next, we compared the energetics associated to the different possible AFD phases in order to evaluate the possible phase transitions. An analysis of the computed and measured Raman active modes for candidate structures was addressed in Sec.~\ref{sec:raman}. In Fig.~\ref{fig:dft_calc}.d, we show the evolution of the thermodynamic potential energy, the enthalpy $H=U+PV$, where $U$ is the DFT-calculated internal energy of the system associated with the relaxed volume $V$ of the AFD-structures under the constraint of given hydrostatic stress ($\sigma_{ij}$), corresponding to the external hydrostatic pressure $P$. We see that there is a continuous increase of the energy gain as a function of pressure for the considered AFD structures with respect to the reference cubic. In particular, the tetragonal $I4/mcm$ phase, to which corresponds the biggest oxygen rotation angle $\Theta$ (Fig.~\ref{fig:dft_calc}.e), is the lowest energy configuration at all studied pressures. Noteworthy, the in-phase ($a^0a^0c^+$) rotation pattern, which define the tetragonal partner $P4/mbm$, produces a much lower gain of energy. In particular, in the compressed cubic cell, it can be seen that,  there is a significant imbalance between the energy gain brought by the condensation of the $c^+$ rotation and by the $c^-$ ones, in favor of the latter, as shown in Fig.~S3.c. This behavior is similar to what also observed in SrTiO$_3$ and PbTiO$_3$~\cite{sharma_2014,sharma_thesis}, and thus it could be a typical feature of perovskite oxides with a tolerance factor $t$ close to $1$. Clearly, the interplay between pressure and strain relaxation stabilize the $a^0a^0c^-$-AFD tetragonal phase with $I4/mcm$ space group at the expense of all other candidates. Moreover, the phonon spectrum of the $I4/mcm$ phase ($a^0a^0c^-$), calculated at \SI{20}{\GPa}, does not display any instability or any mode softening (cf. Table SI and Fig.~S2), confirming that the tetragonal phase is dynamically stable and suggesting that no further phase transition is expected in the investigated pressure range. 

We also considered the possibility for a transition to the ($a^-b^+a^-$) tilt system, corresponding to the very common and stable orthorhombic $Pnma$ structure. For this case, the system relaxes back to the orthorhombic $Imma$ ($a^-b^0a^-$) phase suppressing the $b^+$ rotation, also in line with past studies~\cite{Chen2018, post_post_Perovskite}. Therefore, this highlights that, in BaZrO$_3$, the condensation of one rotational mode suppresses the partner one. It can be surprising at first sight, given how common the $Pnma$ structure is in perovskite systems and the various tilt instabilities appearing at high pressure (e.g. at \SI{60}{\GPa}). 
Nevertheless, a careful analysis of the phonon dispersion and character of the modes at the different high-symmetry $\bm q$-points of the cubic-BZ zone reveals crucial features. In $Pnma$ perovskites such as CaTiO$_3$ and SrZrO$_3$ at \SI{0}{\GPa}~\cite{amoroso2018,Safari_SZO}, it is the competition and coexistence of in-phase and anti-phase rotations together with the trilinear coupling of the latter modes with the A-cation antipolar displacement, in turn also linked to the existence of an unstable polar zone-center mode in the parent cubic phase, mainly driven by the A-cation, to lower the ground state energy~\cite{benedek2013,Miao_2013,Safari_SZO}. Instabilities involving displacements of the A cation is therefore a key ingredient that is missing in BaZrO$_3$ at high pressure, and a fortiori at ambient pressure. At ambient pressure, cubic BaZrO$_3$ displays neither the unstable in-phase tilt mode nor any polar instability. Under pressure, these modes become indeed unstable but (i) the pattern of distortion associated to the unstable $\Gamma$-mode is mainly characterized by a polar motion of the zirconium atoms at the B-site against the oxygen cage; (ii) the phonon branch related to the Ba-displacements stay higher in energy with respect to the CaTiO$_3$ and SrZrO$_3$ cases. As a result, the $Pnma$ phase is never stabilized. 

\section{Conclusion}
In summary, we have reported a combination of Raman spectroscopy, x-ray diffraction and first-principle calculations aimed at clarifying the high-pressure behaviour of BaZrO$_3$. We confirm the picture according to which BaZrO$_3$ undergoes a single phase transition around \SI{10}{\GPa} to a tetragonal $I4/mcm$ phase and retains this structure until the highest pressures investigated -- here \SI{45.1}{\GPa}. We have reconciled this result with previous claims for the existence of a second phase transition, and in particular clarified the Raman mode assignment of the high-pressure phase. From our analysis, it turns out that a second transition to an orthorhombic phase or any other tilted phase is, in fact, not expected. Additionally, we explain why a transition to the common orthorhombic $Pnma$ phase is not expected for this perovskite system. Instead, our study draws a picture where the stability and dominance of the tetragonal phase with its single anti-phase tilt is reinforced with high pressure, with a tilt angle that can reach values as large as 13$^{\circ}$ at \SI{60}{\GPa} according to DFT. The question of its stability limit at even higher pressures therefore remains an open question. According to the calculated phonon dispersion curves, softening of polar modes, and subsequent competition with antiferrodistortive distortions should be expected. In particular, following other recent theoretical studies, two candidate phase transitions can be considered: the first is a transition to a polar phase combining tilts and polar cation displacements; the second is a direct transition to a post-perovskite phase. We anticipate that measurements in the \SI{100}{\GPa} (MBar) range, possibly complemented by in-situ heating to overcome energy barriers, will be necessary to find out which of those two possible options prevail.

\section*{Acknowledgments}
* C.T., D.A. and R.O. contributed equally to this paper and are therefore sharing first authorship. 

The authors are grateful to M.~Mezouar and W.A.~Crichton (ESRF ID27) for in-house beamtime allocation and for their help during Neon gaz loading of the DACs. We acknowledge SOLEIL for provision of synchrotron radiation facilities under proposal N.~20191842. This work was supported by the Innovative Training Networks (ITN) Marie Sklodowska-Curie Actions-European Joint Doctorate in Functional Material Research (EJDFunMat) (Project No. 641640). DFT-based calculations have been performed on the NIC4 and NIC5 clusters hosted at the University of Li\`ege, within the ‘Consortium des \'Equipements de Calcul Intensif’ (C\'ECI), funded by F.R.S-FNRS (Grant No. 2.5020.1) and by the Walloon Region. C.T., M.G., J.K. acknowledge financial support from the Fond National de Recherche Luxembourg through a PEARL Grant (No. FNR/P12/4853155/Kreisel). D.A. is grateful to S. Picozzi (CNR-SPIN) and to M. Verstraete and B. Dup\'e (ULiege) for the time allowed to work on the writing of this paper.

\bibliography{BZO_pressure}

\clearpage



\end{document}


\title{Stability of the tetragonal phase of BaZrO$_3$ under high pressure\\ --\\Supporting information}

\author{Constance Toulouse*}
\affiliation{\UniLu}
\author{Danila Amoroso*}
\affiliation{\ULiege}
\author{Robert Oliva*}
\affiliation{\UniLu}
\author{Cong Xin}
\affiliation{\ICMCB}
\affiliation{\LIST}
\author{Pierre Bouvier}
\affiliation{\Neel}
\author{Pierre Fertey}
\affiliation{\Soleil}
\author{Philippe Veber}
\affiliation{\ICMCB}
\affiliation{\ILM}
\author{Mario Maglione}
\affiliation{\ICMCB}
\author{Philippe Ghosez}
\affiliation{\ULiege}
\author{Jens Kreisel}
\author{Mael Guennou}
\affiliation{\UniLu}


\maketitle

\tableofcontents

\clearpage

\section{Calculation of phonon band structures for cubic BaZrO$_3$ at high pressure}

\begin{figure*}[h]
\begin{center}
\includegraphics[width=\textwidth]{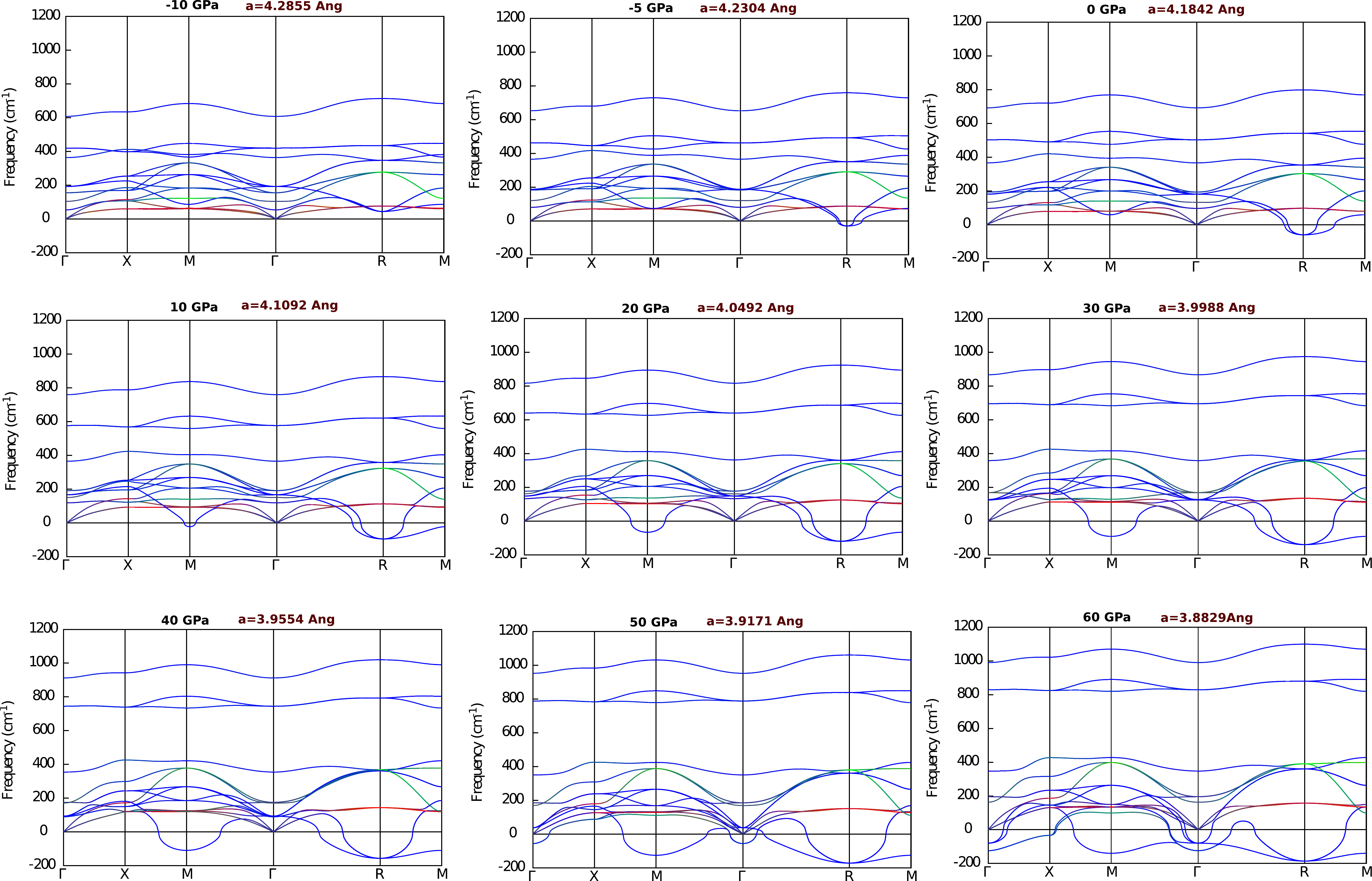} 
\caption{DFPT-calculated phonon dispersion curves in compressed cubic BaZrO$_3$ at several pressures up to \SI{60}{\GPa}. Red,green and blue phononic branches involve mostly Ba, Zr and O vibrations, respectively.   }
\end{center}
\end{figure*}

\clearpage

\section{Calculation of phonon band structures for tetragonal B\lowercase{a}Z\lowercase{r}O$_3$ at high pressure}

\begin{figure*}[h]
\begin{center}
\includegraphics[width=\textwidth]{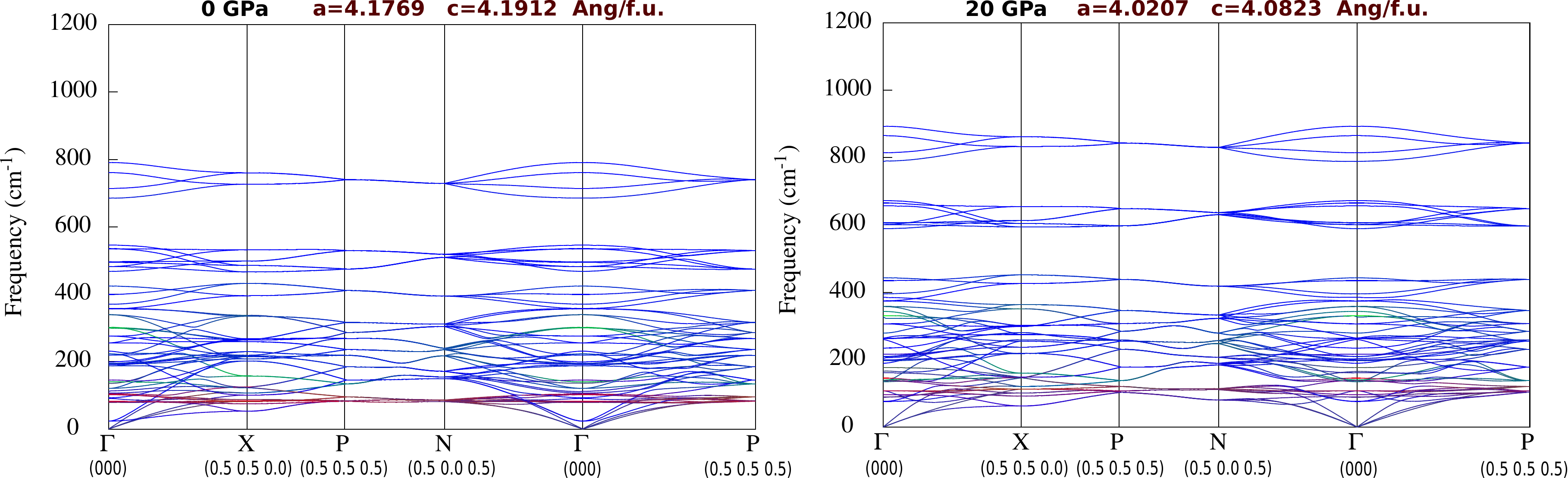} 
\caption{DFPT-calculated phonon dispersion curves in compressed tetragonal $I4/mcm$ phase demonstrating its dynamical stability. Red,green and blue phononic branches involve mostly Ba, Zr and O vibrations, respectively.}
\end{center}
\end{figure*}

\clearpage

\section{Calculated structural properties and internal energies under pressure}

\begin{figure*}[h]
\begin{center}
 \label{fig:exp_structure}
 \vspace{.5cm}
 \includegraphics[width=0.9\textwidth]{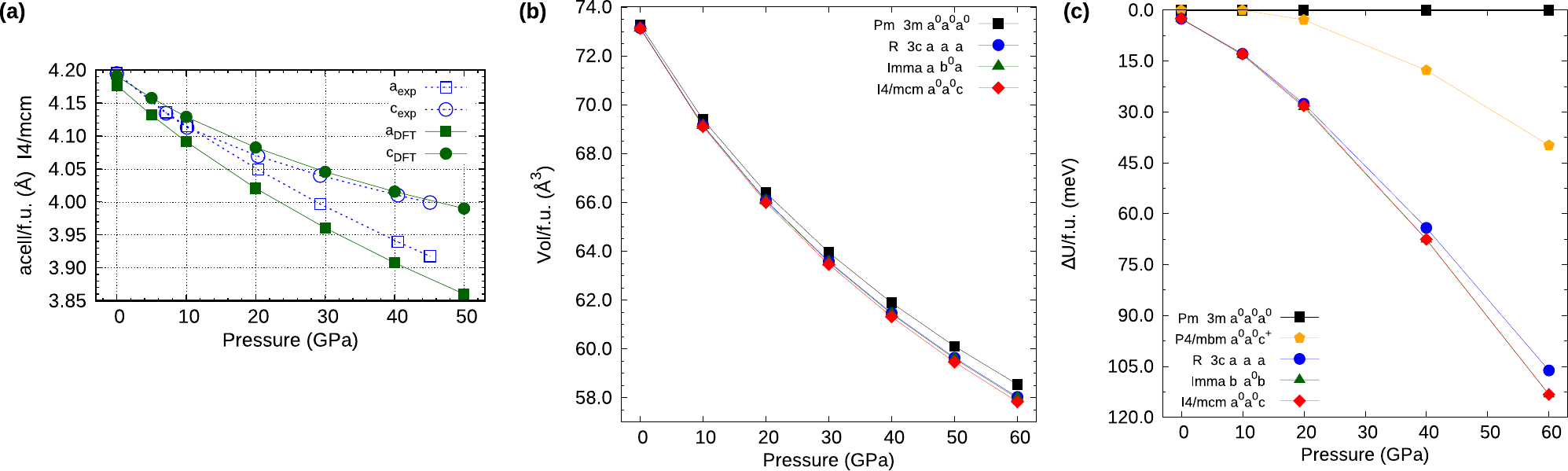} 
 \caption{DFT-calculated {\bf (a)} lattice parameters for the tetragonal $I4/mcm$ phase compared to the Rietvelt refined data, {\bf (b)} relaxed volume for each AFD structure, and {\bf (c)} internal energy ($U$) gain with respect to the high-symmetry cubic phase from relaxed AFD distortions within compressed cubic volume, as a function of hydrostatic pressure.} 
\label{fig:dft_stucture}
\end{center}
\end{figure*}

\clearpage

\section{Reciprocal space maps}

\begin{figure*}[h]
\begin{center}
\includegraphics[width=0.4\textwidth]{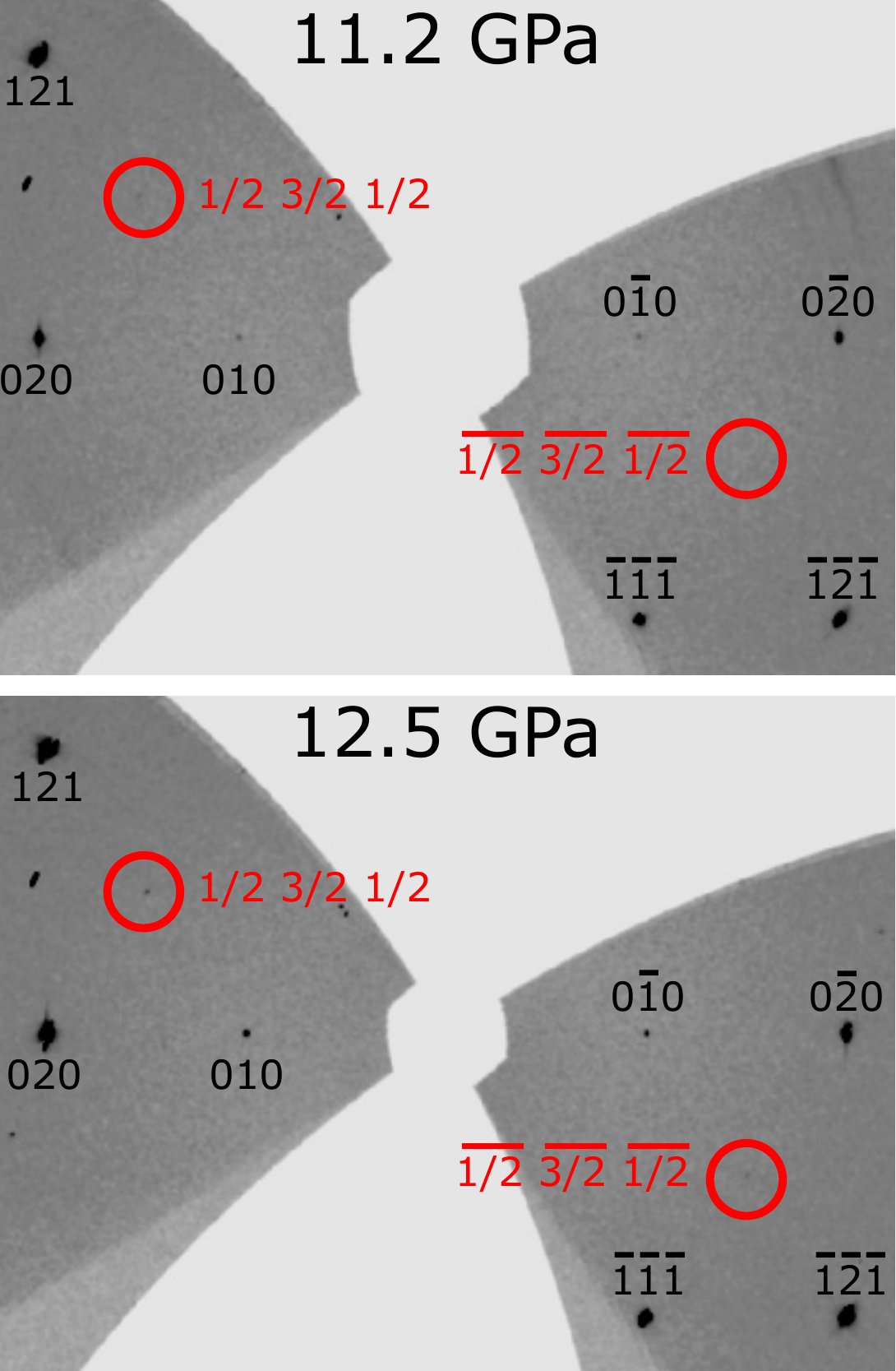} 
\caption{Reciprocal space maps showing the emergence of superstructure reflections in a single crystal diffraction pattern. For anti-phase tilts, the expected reflections have indices of the form $(hkl)/2$ where $h$, $k$, and $l$ are odd and at least one of the indices is different from the others. If the transition is to the tetragonal $I4/mcm$ in particular with the tetragonal axis along $c$, we have $h\neq k$. Further details can be found for example in~\cite{Mitchell2002}. Here, the $(\nicefrac{3}{2}\,\nicefrac{1}{2}\,\nicefrac{1}{2})$ superstructure peaks indicative of antiphase tilting because visible already at \SI{11.2}{\GPa}, although barely above the noise levels. They are much more visible at \SI{12.5}{\GPa}.}
\end{center}
\end{figure*}

\clearpage

\section{Analysis of the XRD peak broadening}

\begin{figure*}[h]
\begin{center}
\includegraphics[width=0.6\textwidth]{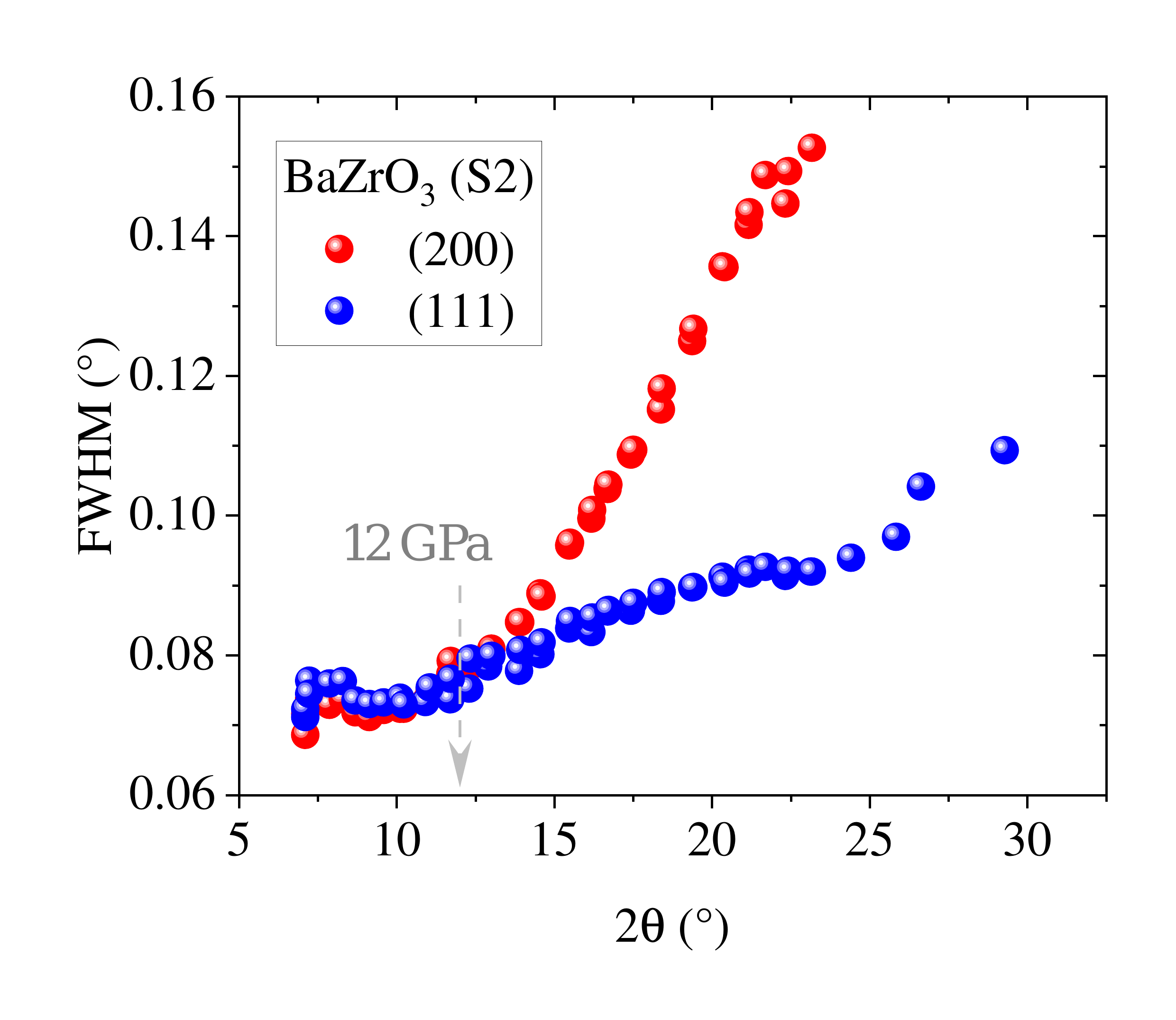} 
\caption{Full width at half maximum of the (200)$_\mathrm{pc}$ and (111)$_\mathrm{pc}$ reflections (in red and blue symbols, respectively) measured for sample S2 at a pressure interval below and above the phase transition (here observed around $9^{\circ}$ and $10.5^{\circ}$, respectively), $P_c$ = 12 GPa. An increased broadening of reflection (200)$_\mathrm{pc}$ with respect reflection (111)$_\mathrm{pc}$ can be observed above the phase transition. This result is consistent with the assignation of the $I4/mcm$ high-pressure phase since reflection (200)$_\mathrm{pc}$ splits into the tetragonal reflections (004)$_\mathrm{t}$ and (220)$_\mathrm{t}$ while reflection (111)$_\mathrm{pc}$ becomes (202)$_\mathrm{t}$ for the tetragonal phase. Moreover, this result allows to discard the $R\bar{3}c$ as a possible high-pressure phase, since a splitting would be expected for (111)$_\mathrm{pc}$ due to a peak splitting and no additional broadening (nor reflection splitting) would be expected for (200)$_\mathrm{pc}$, contrary to observation.}
\end{center}
\end{figure*}

\clearpage

\section{F-\lowercase{f} plots}

\begin{figure*}[h]
\begin{center}
\includegraphics[width=0.6\textwidth]{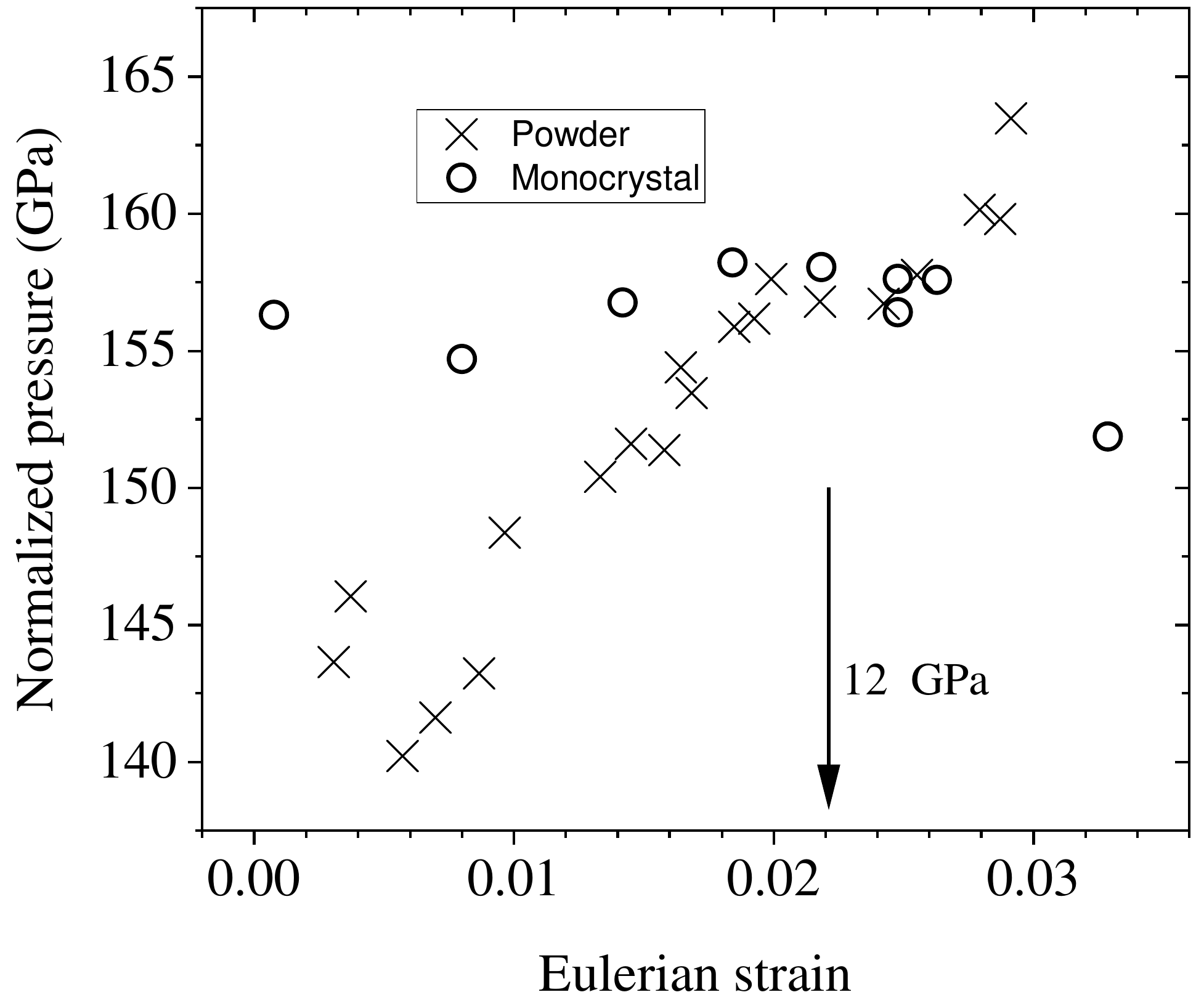} 
\caption{F-f plot for powder and monocrystal XRD high-pressure data. The data lie on a horizontal line for the monocrystal, which evidences that a 2$^\mathrm{nd}$-order Birch–Murnaghan EoS is adequate to describe the $P$-$V$ dependency of BaZrO$_3$ in its cubic phase. However, for the case of our measurements on powder BaZrO$_3$, a significant slope is found, indicating that $K'>4$ and that a 3$^\mathrm{rd}$-order Birch-Murnaghan EoS is required in this case. The difference can be accounted for by the choice of Ne as PTM for the powder measurements vs. He for the single crystal. Helium solidifies at \SI{12.1}{\GPa} and provides in that range better hydrostaticity that Neon that solidifies at 4.8~GPa~\cite{Klotz2009}.  Besides, single crystals are by definition free of intergrain stresses. In the main manuscript we use the 3$^\mathrm{rd}$-order EoS in order to be consistent in the treatment of the powder data for both the low pressure and the high-pressure phases. For the case of our powder data a 2nd order fit yields bulk moduli of 161.3(1.2) and 176.1(5) GPa and volumes-to-zero-pressure of 73.70(2) and 73.38(2) {\AA}$^3$ for the cubic and tetragonal phase, respectively. It would be desirable to perform additional measurements at higher pressure with a single crystal and in helium to revise the parameters for the EoS of BaZrO$_3$.}
\end{center}
\label{Ffplot}
\end{figure*}

\clearpage

\section{Residuals for the fit to the equation of state.}

\begin{figure*}[h]
\begin{center}
\includegraphics[width=0.6\textwidth]{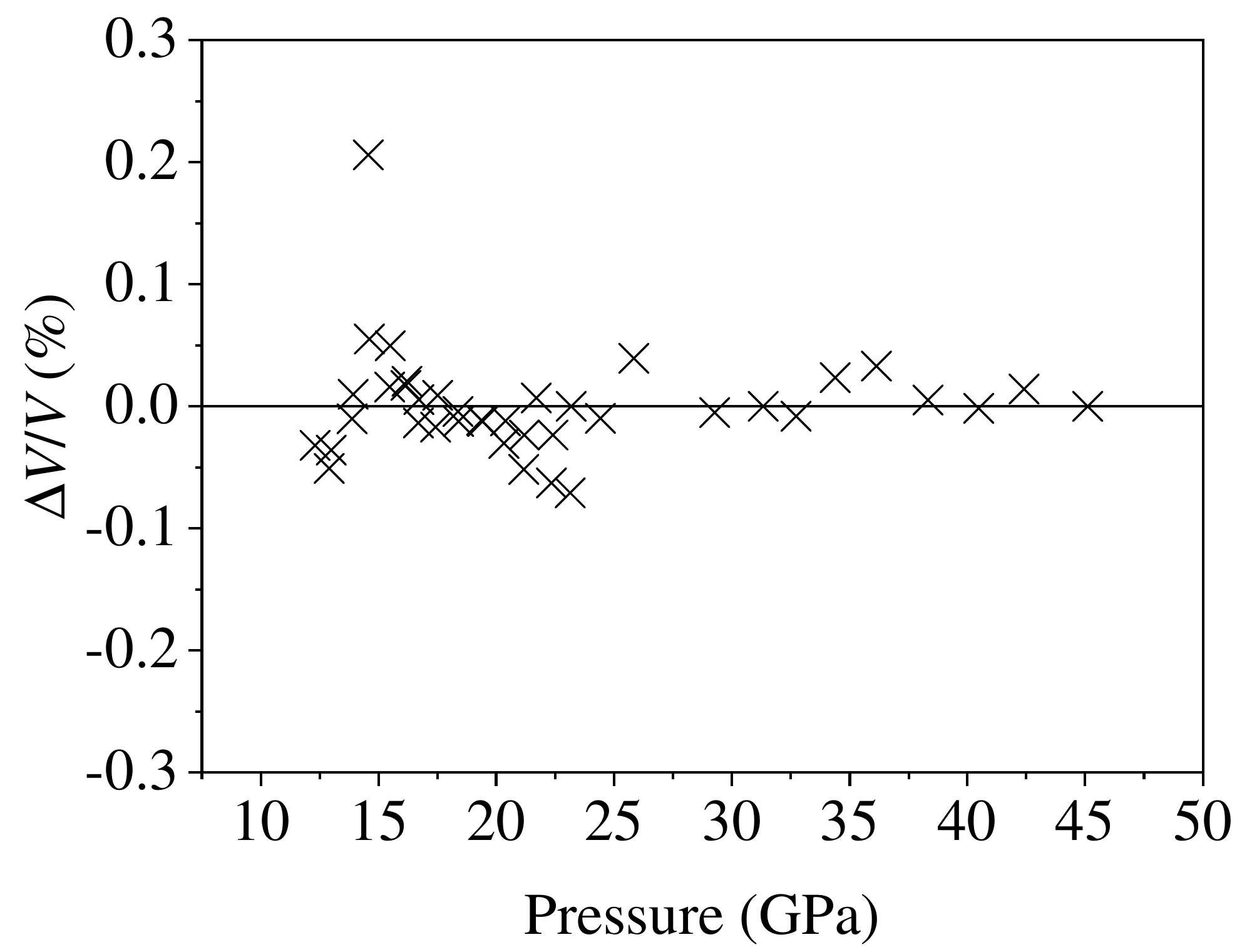} 
\caption{Experimental deviations of powdered XRD measurements from the fitted 2$^\mathrm{nd}$-order Birch–Murnaghan EoS to the $I4/mcm$ structure are smaller than 0.1\% indicating that no additional first-order structural phase transition takes place at least up to \SI{45}{\GPa}. The outlier around \SI{15}{\GPa} is an artifact due to a pressure misscalibration.}
\end{center}
\end{figure*}

\clearpage

\section{Zone center phonon frequencies for the tetragonal phase}

\begin{table}[h]
\centering
\small
\begin{tabular}{ccc|ccc}
\toprule
\multicolumn{6}{c}{\bfseries{$\bm{I4/mcm}$-T}} \\
\multicolumn{3}{c}{0 GPa} & \multicolumn{3}{c}{20 GPa} \\
\cmidrule(lr){1-3} \cmidrule(lr){4-6} 
activity & $\omega$ (cm$^{-1}$) & mode &  activity & $\omega$ (cm$^{-1}$) & mode \\ 
\cmidrule(lr){1-3} \cmidrule(lr){4-6} 
  Raman  & 23.49  & $E_g$    & Raman  & 76.06  & $E_g$    \\
  Raman  & 92.72  & $A_{1g}$ & Raman  & 137.85 & $E_{g}$ \\
  Raman  & 101.82 & $E_g$    & Raman  & 145.29 & $B_{2g}$     \\
  Raman  & 104.36 & $B_{2g}$ &   IR   & 160.76 & $E_u$     \\
   IR    & 106.52 & $E_u$    &   IR   & 176.44 & $A_{2u}$  \\ 
   IR    & 113.38 & $A_{2u}$ & Raman  & 189.47 & $A_{1g}$   \\
   IR    & 191.57 & $E_u$    &   IR   & 196.57 & $E_u$      \\      
   IR    & 194.52 & $A_{2u}$ &   IR   & 208.80 & $A_{2u}$   \\ 
   IR    & 194.72 & $E_u$    &   IR   & 209.55 & $E_u$      \\
  silent & 199.53 & $B_{1u}$ & silent & 234.51 & $B_{1u}$   \\
  silent & 301.13 & $A_{1u}$ & silent & 330.40 & $A_{1u}$   \\
   IR    & 302.03 & $E_u$    &   IR   & 343.30 & $E_u$      \\
  Raman  & 356.41 & $E_g$    & Raman  & 374.37 & $E_g$      \\
  Raman  & 358.05 & $B_{2g}$ & Raman  & 384.56 & $B_{2g}$   \\
   IR    & 494.37 & $E_u$    &   IR   & 605.56 & $A_{2u}$   \\
   IR    & 496.48 & $A_{2u}$ &   IR   & 606.84 &  $E_u$     \\
  silent & 534.22 & $A_{2g}$ & silent & 656.16 & $A_{2g}$   \\
  Raman  & 535.33 & $B_{1g}$ & Raman  & 665.01 &  $B_{1g}$  \\
  silent & 790.83 & $A_{2g}$ & silent & 892.70 & $A_{2g}$   \\  
\bottomrule
\end{tabular}
\caption{Calculated phonon frequencies $\omega$ (in cm$^{-1}$) at the $\Gamma$ (0 0 0) point of the $I4/mcm$ Brillouin zone, for structure relaxed under the constrain of hydrostatic pressure, $0$~GPa and $20$GPa respectively.}
\label{bzo_modes}
\end{table}

\clearpage

\section{Lattice parameters from powder XRD measurements}

\begin{table}[h]
\centering
\small
\begin{tabular}{c|c}
\toprule
\multicolumn{2}{c}{\bfseries{$Pm\bar3m$}} \\
Pressure (GPa) & $a$ ({\AA}) \\
1.34(1)    &	4.1828(1)	\\
1.66(1)    &	4.1800(1)	\\
2.47(1)    &	4.1718(1)	\\
3.07(1)    &	4.1665(1)	\\
3.88(1)    &	4.1597(1)	\\
4.50(1)    &	4.1557(1)	\\
6.42(1)    &	4.1409(1)	\\
7.09(1)    &	4.1361(1)	\\
7.75(1)    &	4.1310(1)	\\
8.25(1)    &	4.1283(1)	\\
8.42(1)    &	4.1268(1)	\\
9.46(1)    &	4.1202(1)	\\
9.92(1)    &	4.1172(1)	\\
10.38(1)	&	4.1146(1)	\\
11.40(1)	&	4.1071(1)	\\
12.82(1)	&	4.0976(1)	\\

\bottomrule
\end{tabular}
\caption{Rietveld-refined lattice parameter of powdered BaZrO$_3$ from x-ray measurements (sample S1).}
\label{fitted_cubic_XRD}
\end{table}

\begin{table}[h]
\centering
\small
\begin{tabular}{c|c|c}
\toprule
\multicolumn{3}{c}{\bfseries{$I4/mcm$}} \\
Pressure (GPa) & $a$ ({\AA})  & $c$ ({\AA}) \\
12.30(1)	&	5.8046(2)	&	8.1932(7)	\\
12.91(1)	&	5.7920(2)	&	8.2082(7)	\\
13.00(1)	&	5.7992(2)	&	8.1831(7)	\\
13.88(1)	&	5.7867(2)	&	8.1848(7)	\\
13.92(1)	&	5.7821(2)	&	8.1943(7)	\\
14.56(1)	&	5.7742(2)	&	8.1784(7)	\\
14.60(1)	&	5.7790(2)	&	8.1762(7)	\\
15.47(1)	&	5.7740(2)	&	8.1629(7)	\\
15.50(1)	&	5.7746(2)	&	8.1576(7)	\\
16.17(1)	&	5.7665(2)	&	8.1600(7)	\\
16.20(1)	&	5.7620(2)	&	8.1715(7)	\\
16.69(1)	&	5.7620(2)	&	8.1575(7)	\\
16.72(1)	&	5.7609(2)	&	8.1580(7)	\\
17.43(1)	&	5.7538(2)	&	8.1557(7)	\\
17.50(1)	&	5.7526(2)	&	8.1552(7)	\\
18.38(1)	&	5.7447(2)	&	8.1496(7)	\\
18.40(1)	&	5.7438(2)	&	8.1519(7)	\\
19.37(1)	&	5.7350(2)	&	8.1448(7)	\\
19.41(1)	&	5.7341(2)	&	8.1465(7)	\\
20.33(1)	&	5.7263(2)	&	8.1400(7)	\\
20.40(1)	&	5.7254(2)	&	8.1391(7)	\\
21.16(1)	&	5.7194(2)	&	8.1348(7)	\\
21.19(1)	&	5.7181(2)	&	8.1353(7)	\\
21.69(1)	&	5.7129(2)	&	8.1316(7)	\\
22.33(1)	&	5.7097(2)	&	8.1262(7)	\\
22.41(1)	&	5.7080(2)	&	8.1261(7)	\\
23.13(1)	&	5.7030(2)	&	8.1212(7)	\\
23.17(1)	&	5.7013(2)	&	8.1196(7)	\\
24.41(1)	&	5.6914(2)	&	8.1112(7)	\\
25.84(1)	&	5.6789(2)	&	8.0999(7)	\\
29.28(1)	&	5.6523(2)	&	8.0799(7)	\\
31.33(1)	&	5.6363(2)	&	8.0682(7)	\\
32.72(1)	&	5.6258(2)	&	8.0616(7)	\\
34.38(1)	&	5.6128(2)	&	8.0522(7)	\\
36.14(1)	&	5.6001(2)	&	8.0417(7)	\\
38.32(1)	&	5.5854(2)	&	8.0298(7)	\\
40.47(1)	&	5.5708(2)	&	8.0195(7)	\\
42.40(1)	&	5.5577(2)	&	8.0090(7)	\\
45.10(1)	&	5.5398(2)	&	7.9983(7)	\\
\bottomrule
\end{tabular}
\caption{Rietveld-refined lattice parameters of powdered BaZrO$_3$ from x-ray measurements (sample S2).}
\label{fitted_tetra_XRD}
\end{table}

\clearpage
\section{Phonon frequencies for the three phases at 20 GPa}

\begin{table*}[h]
	\centering
	\caption{Calculated phonon frequencies $\omega$ (in cm$^{-1}$) at the $\Gamma$ point of the $I4/mcm$, $Imma$, and $R\bar3c$ Brillouin zone under $20$~GPa hydrostatic pressure. The different modes are grouped together based on the symmetry transformations between phases. Raman active modes are indicated in bold and silent modes are in parentheses; the others are infrared active.}
	\label{tab:DFT_modes_freq}
	\setlength{\extrarowheight}{2pt} 
	\begin{tabularx}{\textwidth}{c|l|cc|cc|cc}
		\hline\hline
		Cubic ($Pm\bar{3}m$ - 221) & Related atomic displacements &
		\multicolumn{2}{r|}{Tetra ($I4/mcm$ - 140)} & \multicolumn{2}{r|}{Ortho ($Imma$ - 74)} & \multicolumn{2}{r}{Rhombo ($R\bar{3}c$ - 167)}\\
        \hline
		\multirow{3}{*}{R$_4^+$} & & \textbf{76} & \textbf{\mode{E}{g}} & \textbf{39} & \textbf{\mode{B}{2g}} & \textbf{23} & \textbf{\mode{E}{g}}\\
		 & oxygen octahedra rotations & & & \textbf{69} & \textbf{\mode{B}{1g}} & & \\
		 & & \textbf{189} & \textbf{\mode{A}{1g}} & \textbf{182} & \textbf{\mode{A}{g}} & \textbf{177} & \textbf{\mode{A}{1g}}\\
		\hline
		\multirow{3}{*}{R$_5^+$} & & \textbf{138} & \textbf{\mode{E}{g}} & \textbf{135} & \textbf{\mode{B}{2g}} & \textbf{142} & \textbf{\mode{E}{g}}\\
		 & antiparallel Ba motion & & & \textbf{134} & \textbf{\mode{B}{3g}} & & \\
		 & & \textbf{145} & \textbf{\mode{B}{2g}} & \textbf{139} & \textbf{\mode{A}{g}} & 130 & (\mode{A}{2g})\\
		\hline
		\multirow{3}{*}{$\Gamma_4^-$ (\mode{T}{1u})} & zone center ``Last" modes: & 161 & \mode{E}{u} & 153 & \mode{B}{3u} & 161 & \mode{E}{u}\\
		 & Ba displacement against & & & 162 & \mode{B}{2u} & & \\
		 & ZrO$_6$ octahedra & 176 & \mode{A}{2u} & 171 & \mode{B}{1u} & 148 & \mode{A}{2u}\\
		\hline 
		\multirow{3}{*}{$\Gamma_5^-$ (\mode{T}{2u})} & silent zone-center mode: & 197 & \mode{E}{u} & 180 & \mode{B}{3u} & 184 & \mode{E}{u}\\
		 & antiparallel out-of-plane displace- & & & 209 & \mode{B}{1u} & & \\
		 & -ment of in-plane oxygens & 209 & \mode{A}{2u} & 199 & \mode{A}{u} & 185 & \mode{A}{2u}\\
		\hline
		\multirow{3}{*}{$\Gamma_4^-$ (\mode{T}{1u})} & zone center ``Slater" modes: & 210 & \mode{E}{u} & 196 & \mode{B}{2u} & 211 & \mode{E}{u}\\
		 & out-of-plane displacements & & & 190 & \mode{B}{3u} & & \\
		 & of Zr against in-plane O & 234 & \mode{B}{1u} & 247 & \mode{B}{1u} & 243 & (\mode{A}{1u})\\
		\hline
		\multirow{3}{*}{R$_4^-$} & \multirow{3}{*}{Zr antiparallel displacements} & 330 & (\mode{A}{1u}) & 333 & \mode{A}{u} & 336 & (\mode{A}{1u})\\
		 & & 343 & \mode{E}{u} & 344 & \mode{B}{1u} & 341 & \mode{E}{u}\\
		 & & & & 341 & \mode{B}{2u} & & \\
		\hline
		\multirow{3}{*}{R$_5^+$} & oxygen octahedra & \textbf{374} & \textbf{\mode{E}{g}} & \textbf{363} & \textbf{\mode{B}{2g}} & \textbf{383} & \textbf{\mode{E}{g}}\\
		 & shearing modes & & & \textbf{373} & \textbf{\mode{B}{3g}} & & \\
		 & & \textbf{385} & \textbf{B$_\mathrm{2g}$} & \textbf{390} & \textbf{\mode{A}{g}} & 358 & (\mode{A}{2g})\\
		\hline
		\multirow{3}{*}{$\Gamma_4^-$ (\mode{T}{1u})} & zone center ``Axe" modes: & 606 & \mode{E}{u} & 605 & \mode{B}{3u} & 616 & \mode{E}{u}\\
		 & displacement of apical oxygens & & & 610 & \mode{B}{2u} & & \\
		 &  against in-plane oxygens & 607 & \mode{A}{2u} & 615 & \mode{B}{1u} & 605 & \mode{A}{2u}\\
		\hline
		R$_3^+$  & Jahn-Teller-like distorsions& 656 & (\mode{A}{2g}) & \textbf{662} & \textbf{\mode{B}{1g}} & \textbf{665} & \textbf{\mode{E}{g}}\\
		 & of oxygen octahedra & \textbf{665} & \textbf{\mode{B}{1g}} & \textbf{665} & \textbf{\mode{B}{3g}} & &\\
		\hline
		R$_1^+$ & oxygen octahedra breathing mode & 893 & (\mode{A}{2g}) & \textbf{896} & \textbf{\mode{B}{3g}} & 898 & (\mode{A}{2g})\\
		\hline
		\hline
	\end{tabularx}
\end{table*}

\bibliography{BZO_pressure}